\documentclass[useAMS,usenatbib,12]{article}
\usepackage[dvips]{graphicx}
\usepackage{color}
\title {\textbf{ TRANSISTOR AS A RECTIFIER}}
\author {Raju Baddi\\ National Center for Radio Astrophysics, TIFR, Ganeshkhind, P.O.Bag 3, PUNE 411007.}
\date{}
\begin{document}

\label{firstpage}

\maketitle

\begin{abstract}
Transistor is a three terminal semiconductor device normally used as an amplifier or as a switch. 
Here the alternating current(a.c) rectifying property of the transistor is considered. The ordinary 
silicon diode exhibits a voltage drop of $\sim$0.6V across its terminals. In this article it is shown 
that the transistor can be used to build a diode or rectify low current a.c($\sim$mA) with a voltage 
drop of $\sim$0.03V. This voltage is $\sim$20 times smaller than the silicon diode. The article gives the 
half-wave and full-wave transistor rectifier configurations along with some applications to justify 
their usefulness. \\ 
\end{abstract}

\section{Introduction}
Rectification of a.c is normally carried out using silicon diodes(Kasatkin \& Nemtsov 1986). The 
silicon diode exhibits a voltage drop of $\sim$0.6V across its terminals when forward biased. This 
0.6V is very small voltage if one wants to rectify voltages like 10V or 100V and hence the voltage 
drop across the diode can be neglected. The diode for all practical purposes is considered to conduct 
freely in one direction and completely non-conducting in the other direction(reverse bias).  However 
if one needs to rectify a.c of amplitude 1V, the 0.6V across the silicon diode forms a large 
fraction(60\%). Essentially the silicon diode cannot be used to rectify low voltage a.c($\sim$1V) if 
one is interested in recovering the complete or almost complete(90\%) waveform of the signal.  In this 
article the possibility of using a transistor as a rectifier(Pookaiyaudom et.al) is explored. It is 
seen that the transistor rectifier exhibits a voltage drop of 0.03V or in other words it forms a diode 
with this voltage drop. This voltage is 20 times smaller compared to the silicon diode and 5 times 
smaller compared to the schottky diode(0.15V).  Thus the transistor rectifier can be used to rectify 
a.c as low as 10 times the voltage drop across it, $\sim$300mV.  This article gives half-wave and full-wave 
configurations of the transistor rectifier as shown in Figures 1 and 2 respectively.  \\

\section[]{The Transistor Rectifier}
\begin{figure}[here]
\begin{center}
\includegraphics[width=135mm,height=65mm,angle=0]{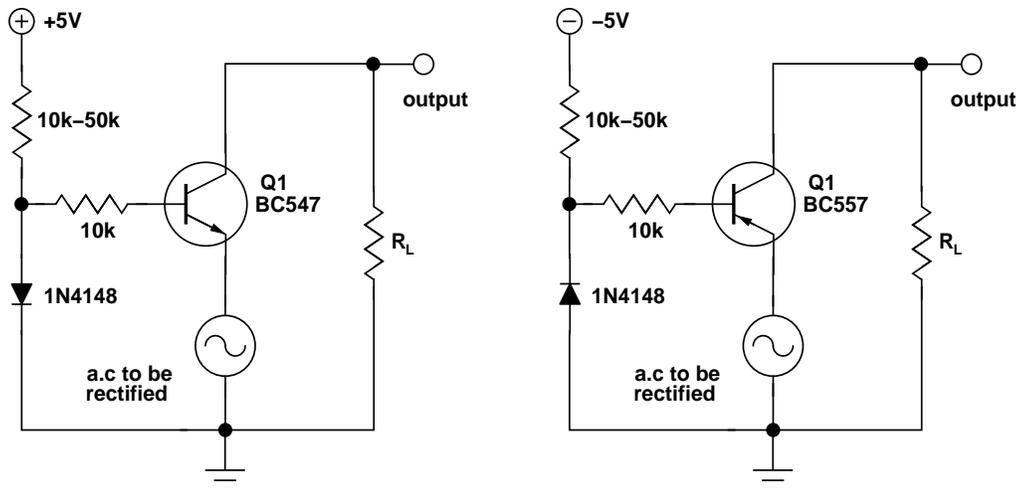}
\caption{Left: NPN version of transistorized a.c rectifier which passes the -ve half cycles. Right: PNP 
version of this circuit would pass the +ve half cycles. The rectified d.c would appear across the load 
resistor R$_L$. Refer Figure 3 for computer simulation results.}
\end{center}
\end{figure}
\begin{figure}[here]
\begin{center}
\includegraphics[width=80mm,height=70mm,angle=0]{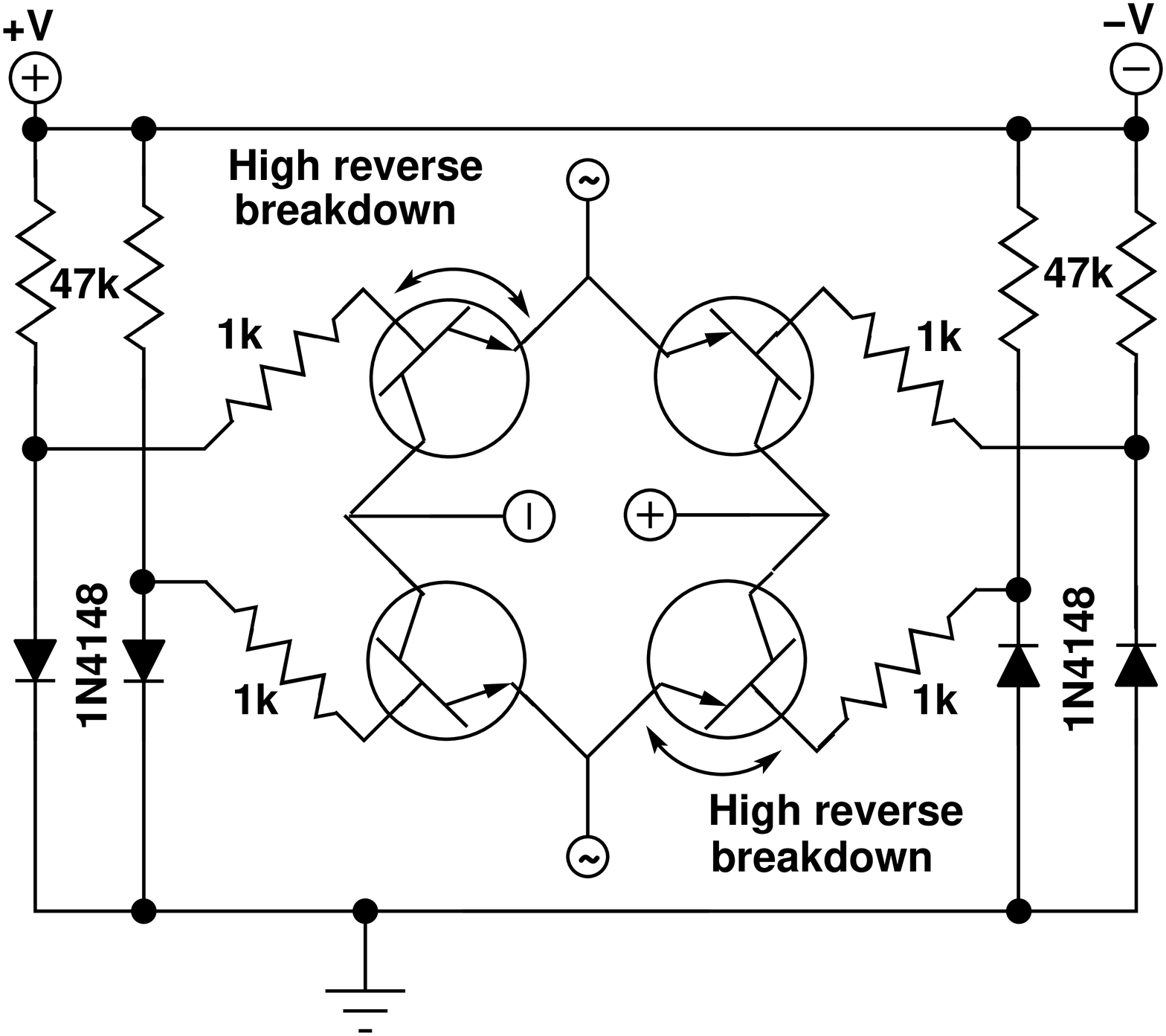}
\caption{A transistorized bridge rectifier using 2-NPN and 2-PNP transistors replacing the diodes in the 
standard configuration. The biasing to the transistors is obtained from the voltage drop across the diodes
(1N4148) in the side arms. The transistors should have high reverse breakdown voltage for BE junction. The 
supply voltage($\pm$V) should be less than this reverse breakdown voltage.}
\end{center}
\end{figure}
\begin{figure}[here]
\begin{center}
\includegraphics[trim = 0cm 1cm 1cm 7.2cm,clip,width=61mm,angle=-90]{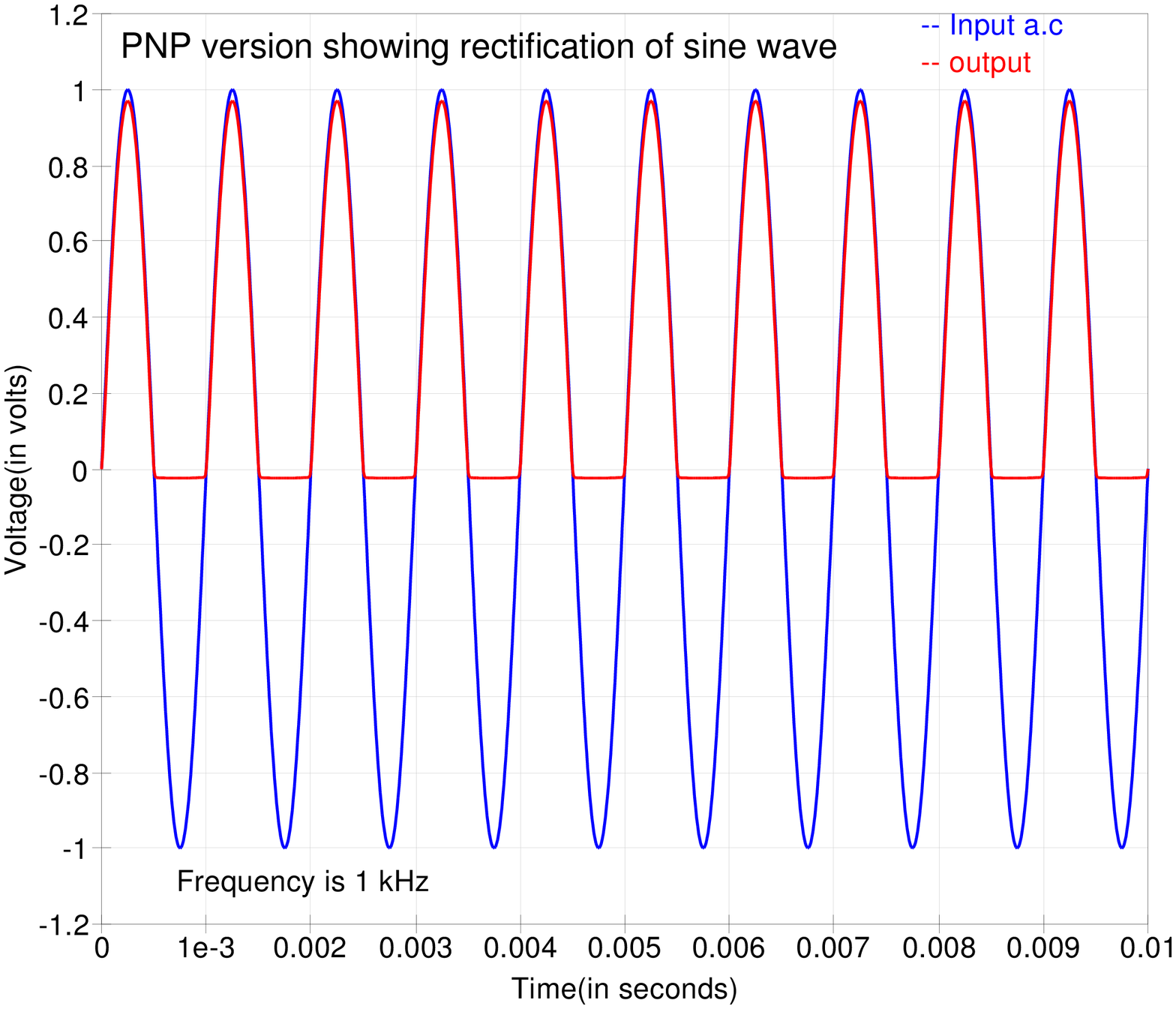} \hspace {5mm}
\includegraphics[trim = 0cm 1cm 1cm 5.0cm,clip,width=61mm,angle=-90]{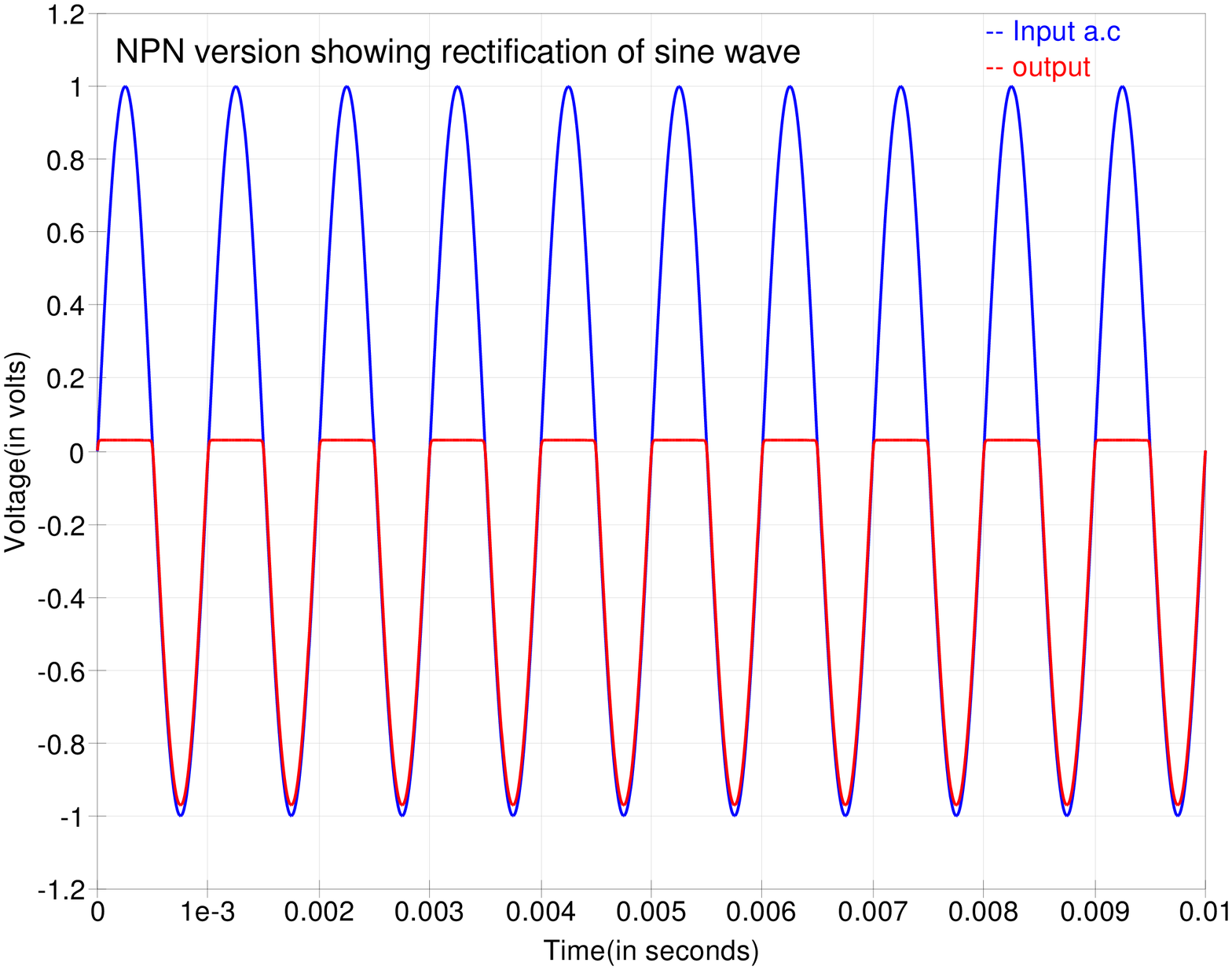}
\caption{Transistorized Rectifier Simulations: Half wave rectifier with a 1V pk-pk and 1kHz 
ac input. Left: PNP version which rectifies positive half cycle. Right: NPN version 
that rectifies negative half cycle. Simulations in linux QUCs.}
\end{center}
\end{figure}

Rectification of alternating current with a single transistor has been considered. This simple circuit 
consisting of a transistor, a biasing diode and a few resistors can rectify a.c as low as 0.3V or even 
smaller. The rectification can be for positive half cycle or negative half cycle depending on the type of 
transistor(pnp/npn respectively) chosen. The schematic circuit of the rectifier is as shown in Figure-1.  
The quiescent forward biasing(Kasatkin \& Nemtsov 1986) of the base-emitter junction of the transistors 
is taken care by the voltage drop across the silicon diodes(1N4148) in the side arms.  The base-emitter junction of the 
silicon transistor and the silicon diode being similar the voltage drops across them are also more or 
less equal. The a.c to be rectified is introduced in the emitter circuit as shown in the Figure-1.  Under 
these conditions any low negative voltage(NPN-version) appearing at the emitter would 
set up an emitter current which would mostly flow to the collector. However a small voltage drop($\sim$0.03V)
across collector and emitter of the transistor is required to maintain this current.  Essentially 
any additional voltage in the emitter circuit other than the balanced out voltage drops of the silicon 
diode and the base-emitter junction will be transfered to the collector circuit dropping across the load R$_L$, as most of the emitter 
current flows to the collector.  The Figure-1 shows both types of rectifiers which can rectify either 
the negative phase or the positive phase of the a.c signal. A full wave bridge rectifier using four 
transistors is shown in Figure-2.  The results of computer simulation are shown in Figure-3. 
During the opposite phase i.e positive voltage w.r.t ground for NPN-version the base-emitter junction 
would be reverse biased however the base-collector junction is forward biased with the voltage drop 
across the biasing diode which is insufficient to setup a current through R$_L$. Thus the circuit behaves 
as a rectifier passing only one phase of the a.c. The following appendix gives analysis and applications 
of the transistor rectifier.

\appendix
\section[]{Analysis of the Rectifier}
\begin{figure}[here]
\begin{center}
\includegraphics[width=110mm,height=55mm,angle=0]{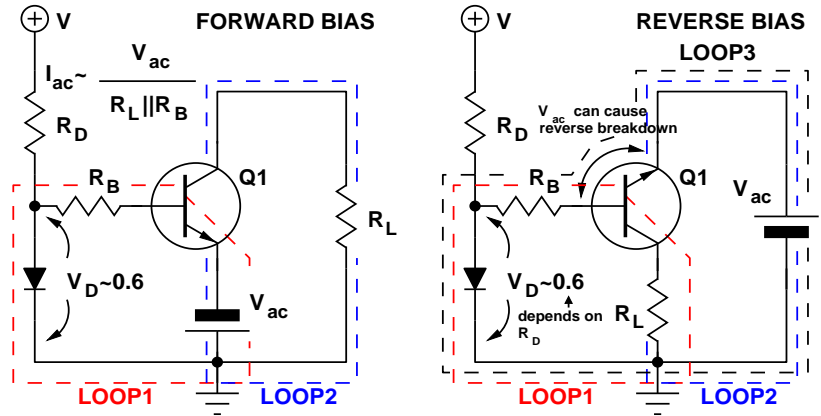}
\caption{Forward(left) and Reverse(right) bias configurations of the transistor rectifier.  
During forward bias V$_{ac}$ sees loops 1 and 2 and current flows in both the loop resistors. Essentially 
V$_{ac}$ sees an impedance of R$_B||$R$_L$(when R$_D$ is sufficiently low, else impedance is higher).  If R$_B$ 
is considerably higher than R$_L$ then impedance is $\sim$R$_L$. Under reverse bias the configuration is 
similar to a voltage follower circuit but with emitter in place of collector and viceversa. As 
the voltage drop across the biasing diode is just sufficient for the BC junction only a weak 
constant current flows through R$_L$ which can be seen in Figure 3 as a weak leakage during the cutoff 
phase of the a.c.}
\end{center}
\end{figure}
The forward bias and reverse bias operation of the NPN transistor rectifier are understood as shown in Figure 4. 
During forward bias(left figure) the voltage drop across the biasing diode V$_D\sim$0.6V supplies the necessary base-emitter 
voltage drop for Q1. The negative voltage V$_{ac}$ w.r.t ground during this phase sees the LOOP1 and a current $\sim$V$_{ac}$/R$_B$ 
should flow through R$_B$. This causes a voltage drop of $\sim$V$_{ac}$ across R$_B$. It is assumed that V$_{ac}$/R$_B$ is drawn from the 
biasing diode's(assuming R$_D$ is sufficiently low, however if R$_D$ is high then current through R$_B$ can be lower) current 
which can only cause a small change in its voltage drop as per its I-V characteristics. V$_{ac}$ also sets up current 
V$_{ac}$/R$_L$ in the load resistor. If V$_{ac}$ is further increased it can even lead to reversing the voltage drop across the 
biasing diode(impedance : V$_{ac}$(R$_D$+R$_B$)/(V$_{ac}$+V) $||$ R$_L$). However the upper limit on V$_{ac}$ is set by the reverse break down 
of BE junction (loop 3)which would be unfavourable for the life of the transistor.  Further the value of R$_D$ plays a 
role in the reverse leakage current which is seen as a small constant voltage during the cutoff phase of ac(Figure 3).  
Smaller the value of R$_D$ higher is the voltage drop across biasing diode and more will be the leakage. The same is 
true for R$_B$ as well, however R$_B$ also plays a role in impedance seen by V$_{ac}$. During reverse bias(Figure 4, right) 
the configuration is similar to that of a voltage follower however with the emitter replacing the collector and 
viceversa. Under this condition the voltage drop across the biasing diode is just sufficient to supply the BC junction 
forward biasing and not much is left for the voltage drop across R$_L$ if a current starts flowing through it. However 
a weak current can still flow through R$_L$ which depends on the voltage drop across the diode and in turn on the value 
of R$_D$. This leakage is seen in Figure 3 as a small constant voltage $\sim$few 10mV. The following sections give 
applications of the transistor rectifier.

\subsection[]{Small signal AC voltmeter}
\begin{figure}[here]
\begin{center}
\includegraphics[width=60mm,height=70mm,angle=0]{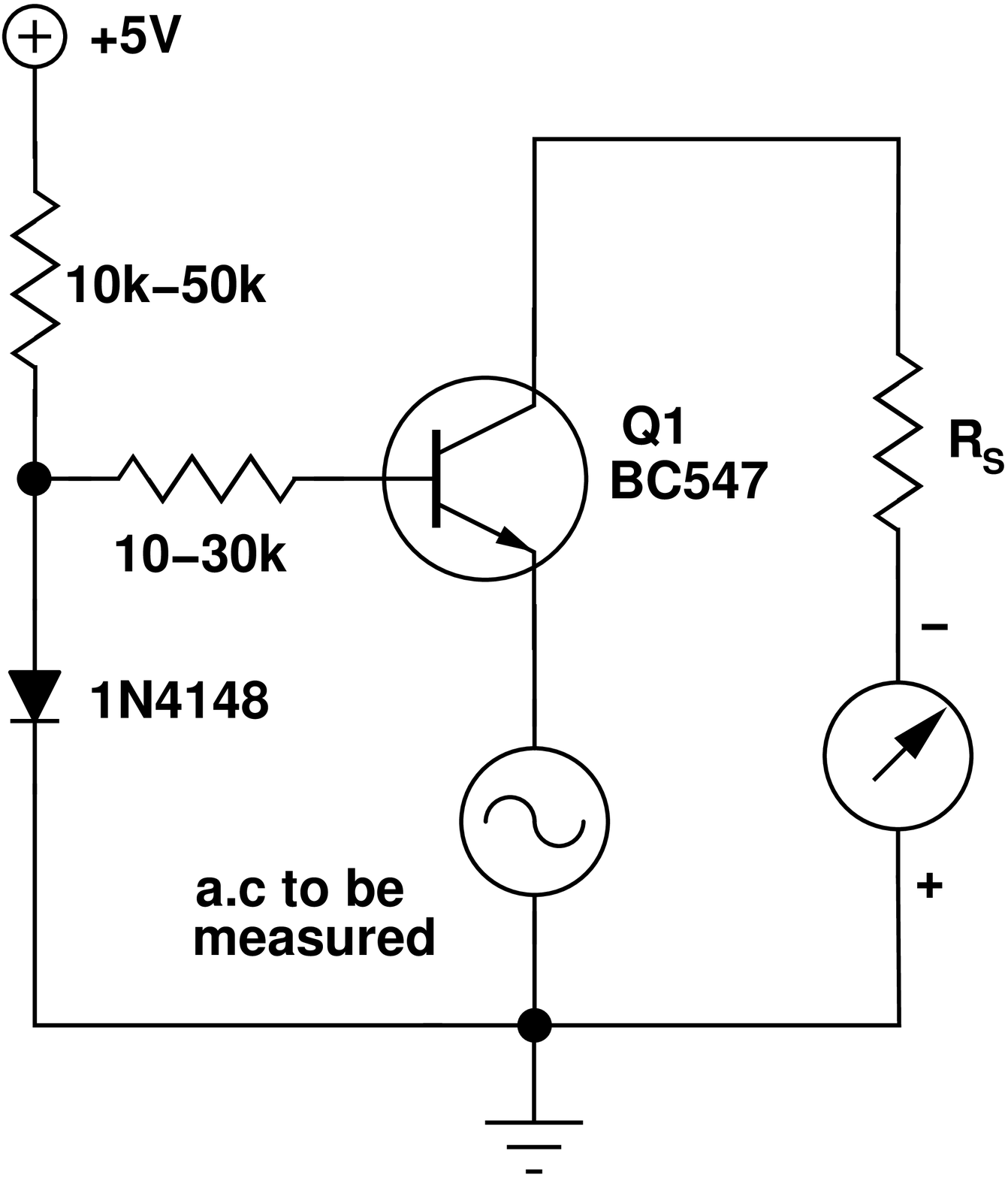} \hspace {3mm}
\includegraphics[width=80mm,height=95mm,angle=0]{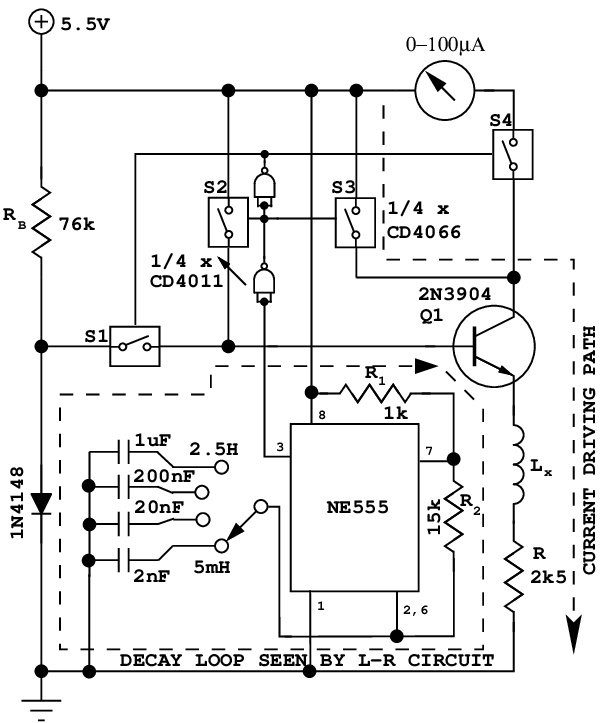}
\caption{Left: Small signal a.c voltmeter. Resistor R$_S$ is chosen such that full scale deflection corresponds to choosen range 
of the instrument. R$_S$ can be either chosen by trial and error or calculated assuming the transistor to be an ideal rectifier.
Right: Inductance Meter using the half wave transistor rectifier(Copyright $\copyright$ 2013 UBM. Reprinted with permission.). R$_2$
has to be adjusted to calibrate the instrument. During transistor on time t$_{on}$ switches S2 and S3 act in parallel reducing the 
effective switch resistance by a factor of 2. The switch resistance can also be reduced by operating the circuit at a higher voltage, 
say 10V or more [5].}
\end{center}
\end{figure}
The half wave rectifier in Figure 1 can be used to build a simple a.c voltmeter(Figure 6) which can measure a.c voltages 
as low as 0.3V which using conventional 600mV voltage drop silicon diode is not possible, however one can still use 
a schottky diode($\sim$150mV). The transistor gives a much smaller(5 times) voltage drop across its CE terminals and acts 
as a diode with $\sim$30mV voltage drop. The schematic circuit diagram of the a.c voltmeter is as shown in Figure 5. \\

\subsection[]{Inductance meter}
The half wave transistor rectifier configuration can be used to build a L-R based inductance meter as shown in Figure 5 right.
It is basically a modification of the half wave rectifier which includes 
appropriate electronic switching to drive the L-R circuit to a certain current and then allow it to decay unhindered through the 
transistor rectifier, however simultaneously measuring the current in the collector circuit.
The transistor has the property of transfering a current from a lower resistance circuit(base-emitter) 
into a higher resistance circuit(base-collector). This property has been exploited here to measure inductance.  
By using a series L-R circuit in the emitter in place of the a.c voltage source it is possible to let a current carrying L-R circuit 
decay completely. At the same time this current can be measured without hindering the decay process in the L-R circuit. This is sufficiently warranted 
by the property of the transistor, for which it also derives its name! Transient analysis(Figure 6) 
of L-R circuit shows that, if in a certain time t$_{off}$ the L-R circuit's current reduces to sufficiently low 
value say 5\% or even less then for a period t$_{off}$ the average current is directly proportional to 
the value of the inductance. This is justified as, higher the value of inductance higher is the energy 
stored(=0.5LI$^2$), for a given current.  We can as well include another constant t$_{on}$ with t$_{off}$ without 
any loss of proportionality.  During t$_{on}$ (transistor on time) the inductor is driven to a maximum current i$_0$$\sim$(V$_S$-0.6)/R 
using proper switching circuitry. The plan to measure inductance is to first drive the L-R circuit to a maximum 
current i$_0$ during the time t$_{on}$. For the period t$_{on}$ the current through the meter is cutoff using the 
switch S4. Switch S3 opens to the +ve of the supply, S2 biases the transistor by connecting to +ve 
of the supply and S1 is off. The current through the L-R circuit evolves as a standard L-R circuit 
connected to a source of voltage V$_S$-0.6 and at the end of t$_{on}$ would have almost reached the maximum 
allowed current through the resistor R, i.e i$_0$. These waveforms have been shown in detail in Figure 6. The 
periods t$_{on}$ \& t$_{off}$ (transistor rectifier time) should be kept as close as possible (i.e equal).  During t$_{off}$ the 
switches S2 and S3 are off and S1 and S4 are on. The current in the inductor now decays through 
the half wave rectifier. Since the transistor half wave rectifier is near to an ideal rectifier the decay 
loop appears essentially as an unhindered closed path to the L-R circuit. A tiny 
portion of the L-R current flows through the base but does not affect the inductance measurement 
seriously due to its negligible magnitude. Majority of the emitter current flows to the collector 
through the meter. As has been shown in the following discussion the average value of this is proportional to 
the inductance L. At the end of t$_{off}$ the current through the inductor would have decayed to 
almost zero value. The end result of the analysis of the inductance meter is contained in the following equation. 
\begin{equation}
L_{actual} = L_{read}\left(\frac{2.5 + R_{coil}}{2.5}\right)^2
\end{equation}

Where R$_{coil}$(in k$\Omega$) is the coil  resistance of the inductor L$_X$ under measurement. This has to be 
measured using a resistance meter and is seriously applicable if L$_X$ has appreciable coil-resistance. L$_{read}$ is the 
value indicated by the inductance meter. The following discussion considers t$_{on}$ and t$_{off}$ phases and quantifies 
various transients and derives the above equation.\\

First the evolution of charging current through the L-R circuit is considered. The magnitude of current at 
a time t in the L-R circuit when a potential difference exists across it is,

\begin{equation}
i = i_0 \left( 1 - e^{-\frac{R}{L_X} t} \right)
\end{equation}

where i$_0$($\sim$ (V$_S$ - 0.6)/R )is the maximum current acheived at time t = $\infty$.  By considering 
the values of the components associated with NE555 we see that t$_{on}$ $\sim$ 20$\mu$s when set for measuring 
a maximum inductance of 5mH. For this setting the maximum value of R/L that is possible is 
2500/0.005 = 500000. So we see that at t = t$_{on}$ the exponential term in (1) is almost 0 as 
(Rt$_{on}$/L$_X$) = 10. This means i = i$_0$ at t = t$_{on}$.  

\begin{figure}[here]
\begin{center}
\hspace{1cm}
\includegraphics[trim = 0cm 4cm 2cm 7cm, clip, width=90mm,angle=0]{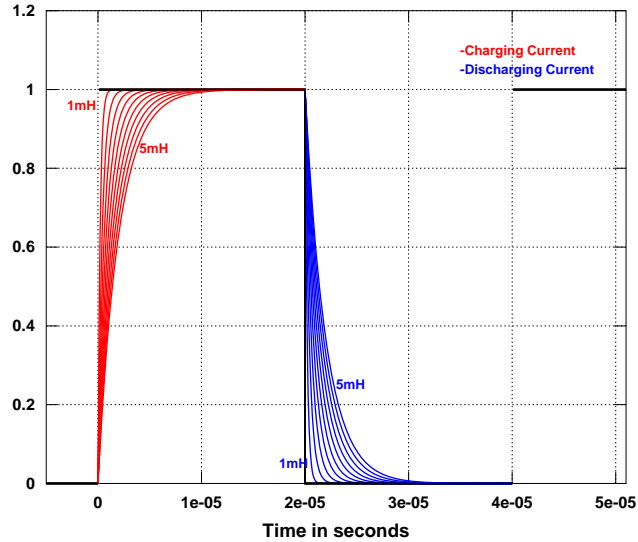}
\caption{Plot of fractional current Vs time for different values of inductor charged through a resistor of 
2.5k$\Omega$. Here decaying current has been called discharging current.}
\end{center}
\end{figure}

	Next we consider the decay or discharge of current during t$_{off}$ . The current at time t in the L-R circuit 
once it starts decay is given by, 
				
\begin{equation}
i = i_0 e^{-\frac{R}{L_X} t}
\end{equation}

Since t$_{off}$ $\sim$ t$_{on}$ we see from the previous argument that i = 0 at t = t$_{off}$.  The plots of evolution/decay 
of currents in the L-R circuit have been shown in Figure 6.\\

The timing diagram(Figure 6) clearly indicates that the current at the end of t$_{on}$ would have reached the 
maximum value or it would have decayed to 0 at the end of t$_{off}$. So the average current through the meter assuming 
that the emitter current almost flows to the collector can be calculated as,

\begin{equation}
i_{avg} = \frac{i_0}{t_{on} + t_{off}} \int_0^{t_{off}} e^{-\frac{R}{L_X} t} dt
\end{equation}

\begin{equation}
i_{avg} = \frac{i_0 L_X}{R \left(t_{on} + t_{off} \right)}
\end{equation}

We see that the average current is directly proportional to the value of the inductance L$_X$ . Since t$_{on}$ $\sim$ t$_{off}$ 
by the choice of R$_1$ and R$_2$ for NE555 astable [6], to measure values of inductances over a wide range  
it is sufficient to change the timing capacitor associated with NE555 as this would proportionally increase (t$_{on}$ + t$_{off}$) leaving 
i$_{avg}$ unaltered. 
To double the range of inductance measurement the value of this capacitor has to be simply doubled. Figure 5 shows the switching 
arrangement that changes the range in steps of decades. \\

	The inductor will also have an associated resistance with it which depends on the length/thickness of the wire. This 
can be accounted by modifying (4) as,

\begin{equation}
i_{avg} = \frac{\left(i_0 R /(R + R_{Lx}) \right) L_X}{R (R + R_{Lx})(t_{on} + t_{off})} R
\end{equation}

\begin{equation}
i_{avg} = \frac{i_0 L_X}{R \left(t_{on} + t_{off} \right)} \left(\frac{R}{R + R_{Lx}} \right)^2
\end{equation}

which is essentially multiplying the  actual inductance by the factor (R/(R+R$_{Lx}$))$^2$ . The manipulation 
in (5) is that due to the introduction of extra resistance R$_{Lx}$  with L$_X$ the peak current to which the 
inductor is driven has reduced from i$_0$ to i$_0$R/(R+R$_{Lx}$). Next in the denominator R has to be replaced 
by (R+R$_{Lx}$). To rewrite and retain the form of (5) we throw an extra R in the numerator and denominator 
and rearrange into (6).   R$_{Lx}$     can be  measured separately using a resistance meter.  R is known to 
be 2.5k$\Omega$ as per this design.  It should be noted that R can be changed to 5k$\Omega$  resulting in 
the modified range of 0-10mH instead of 0-5mH. The  current evolution plots would still be the same with 5mH curve 
indicating the one for 10mH, but the peak current i$_0$ would be  halved. Further R may have to be 
calibrated slightly to get the right value of inductance.  The actual inductance is obtained as the  
observed inductance multiplied by the inverse of the above mentioned factor i.e, ((R+R$_{Lx}$)/R)$^2$. For 
example if R=2.5k$\Omega$ as in our case we have, 
                                                                                                                    
\begin{equation}
L_{actual} = L_{read} \left( \frac{2.5 + R_{Lx}}{2.5} \right)^2
\end{equation}

where R$_{Lx}$ is in k$\Omega$. L$_{actual}$ is the true inductance of the test inductor and 
L$_{read}$ is that shown by the meter. Figure 7 shows a circuit layout to help the reader build the instrument. 
Actual performance of the instrument can be seen at http://youtu.be/Vyd84HvYhyo and http://youtu.be/E7cgomOn5pQ\\

\begin{figure}[h]
\begin{center}
\includegraphics[width=140mm,height=60mm,angle=0]{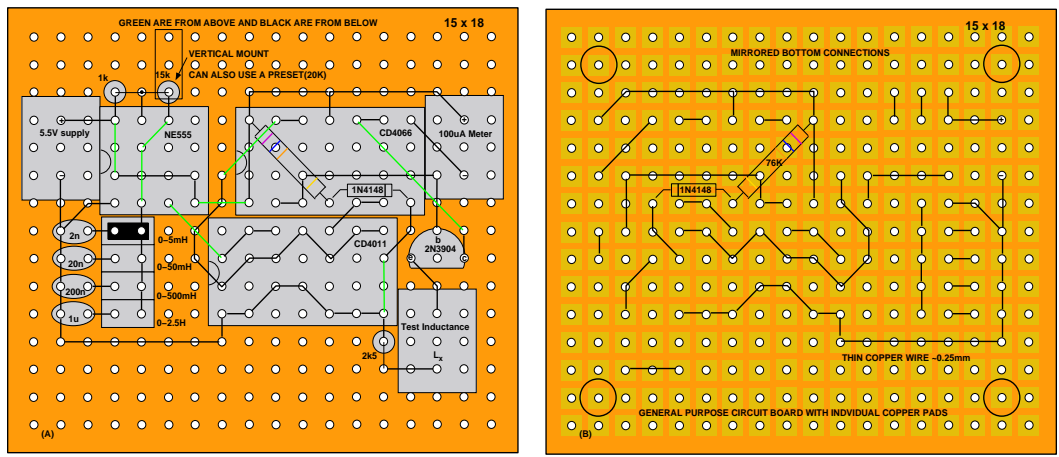}
\caption{Circuit layout for the Inductance Meter.}
\end{center}
\end{figure}

\label{lastpage}

\end{document}